\documentclass{moriondNOTITLES}

\usepackage{myphi}
\newcommand{\sNNn}[1]{\mbox{\sNN $=$ \SI{#1}{\GeV}}}
\graphicspath{{figs/}}

\newcommand{\Photo}{}

\begin{document}
\vspace*{4cm}
\title{News from the strong interaction program of \NASixtyOne}

\author{Antoni Marcinek for the \NASixtyOne Collaboration}

\address{Institute of Nuclear Physics, Polish Academy of Sciences,
Radzikowskiego 152,\\ 31-342 Kraków, Poland}

\maketitle\abstracts{
The NA61/SHINE experiment at the CERN SPS is a multipurpose fixed-target
spectrometer for charged and neutral hadron measurements. Its research program
includes studies of strong interactions as well as reference measurements for
neutrino and cosmic-ray physics. This contribution summarizes a selected set of
new results from our strong interaction program, including system-size and
energy dependence of charged kaon to pion production, \phiM spectra in \ArSc
collisions and enhancement over \pp reactions, femtoscopy of like-sign pions in
\ArSc and \BeBe collisions, $D^0+\overline{D}^0$ (open charm) yield in \XeLa
interactions, and finally an unexpected excess of charged over neutral kaon
production indicating a strong violation of isospin symmetry in \ArSc collisions.
}

\section{Introduction}
The \NASixtyOne experiment has a physics program consisting of
studies of strong interactions, as well as precision measurements for neutrino
and cosmic-ray physics. This contribution focuses on the strong interaction
program which includes the study of the onset of
deconfinement~\cite{Gazdzicki:2010iv}, search for the critical point of strongly
interacting matter,
measurements of open charm and, unexpectedly, isospin-symmetry violation in
multiparticle production.
\par
The \NASixtyOne detector is a multipurpose fixed-target spectrometer at the CERN
SPS. Its main components
are large-volume Time Projection Chambers, two of them immersed in a magnetic
field perpendicular to the beam. This gives \NASixtyOne a significant advantage
over collider experiments --- acceptance covering nearly entire forward
hemisphere, down to $\pt=0$ for charged hadrons.
The detector underwent a major
upgrade during CERN LS2 increasing the data taking rate ten-fold to the order of
\SI{1}{\kHz}, enlarging acceptance of the Vertex Detector, exchanging Time of
Flight walls and adding new forward calorimeter.

\section{Study of the onset of deconfinement and search for the critical point}
\begin{figure}[tb]
  \centering
  \hspace*{-2ex}%
  \includegraphics[height=0.198\textheight,clip, trim=0 10mm 0 0]{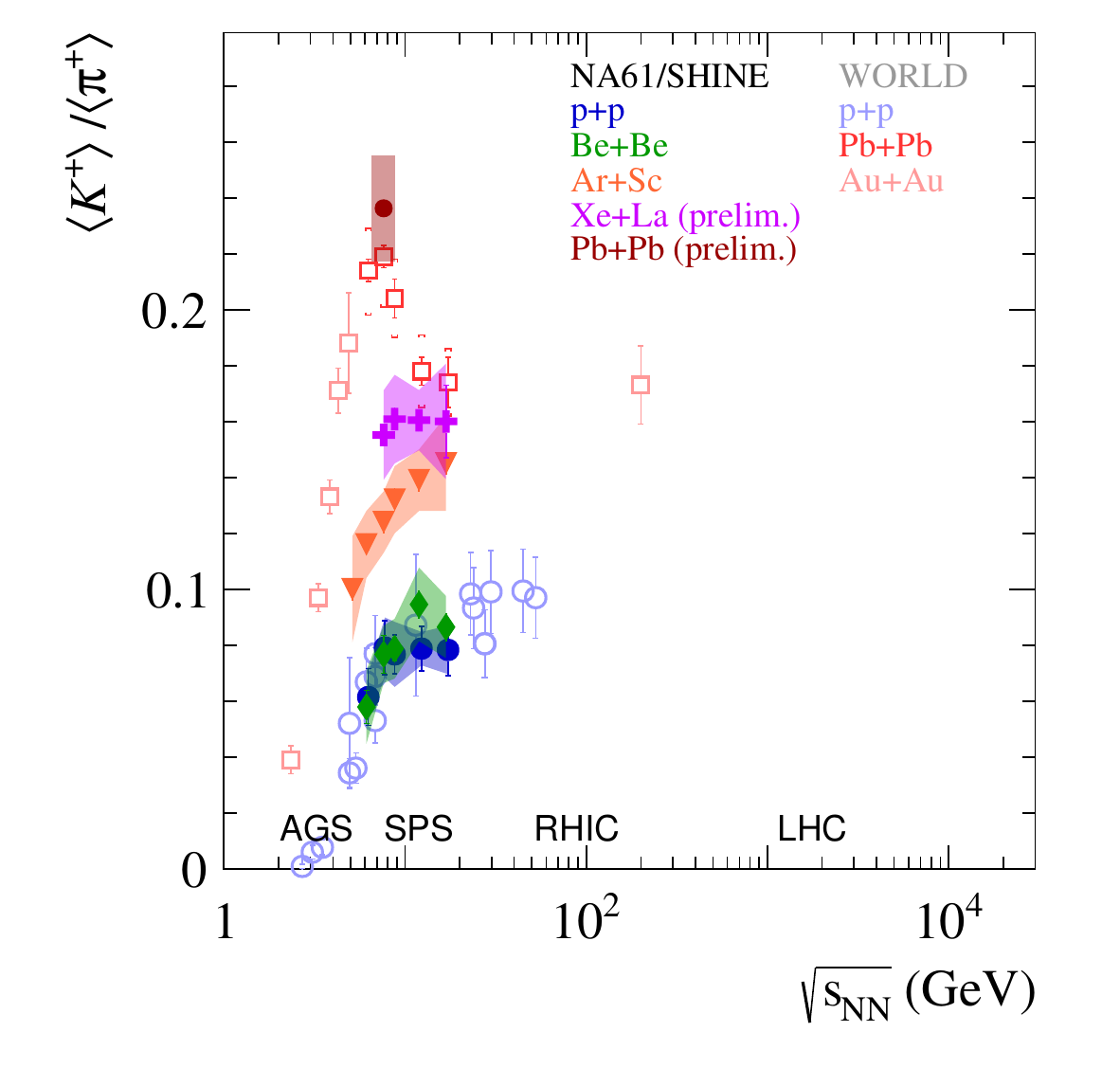}%
  \includegraphics[height=0.198\textheight,clip, trim=0 10mm 0 0]{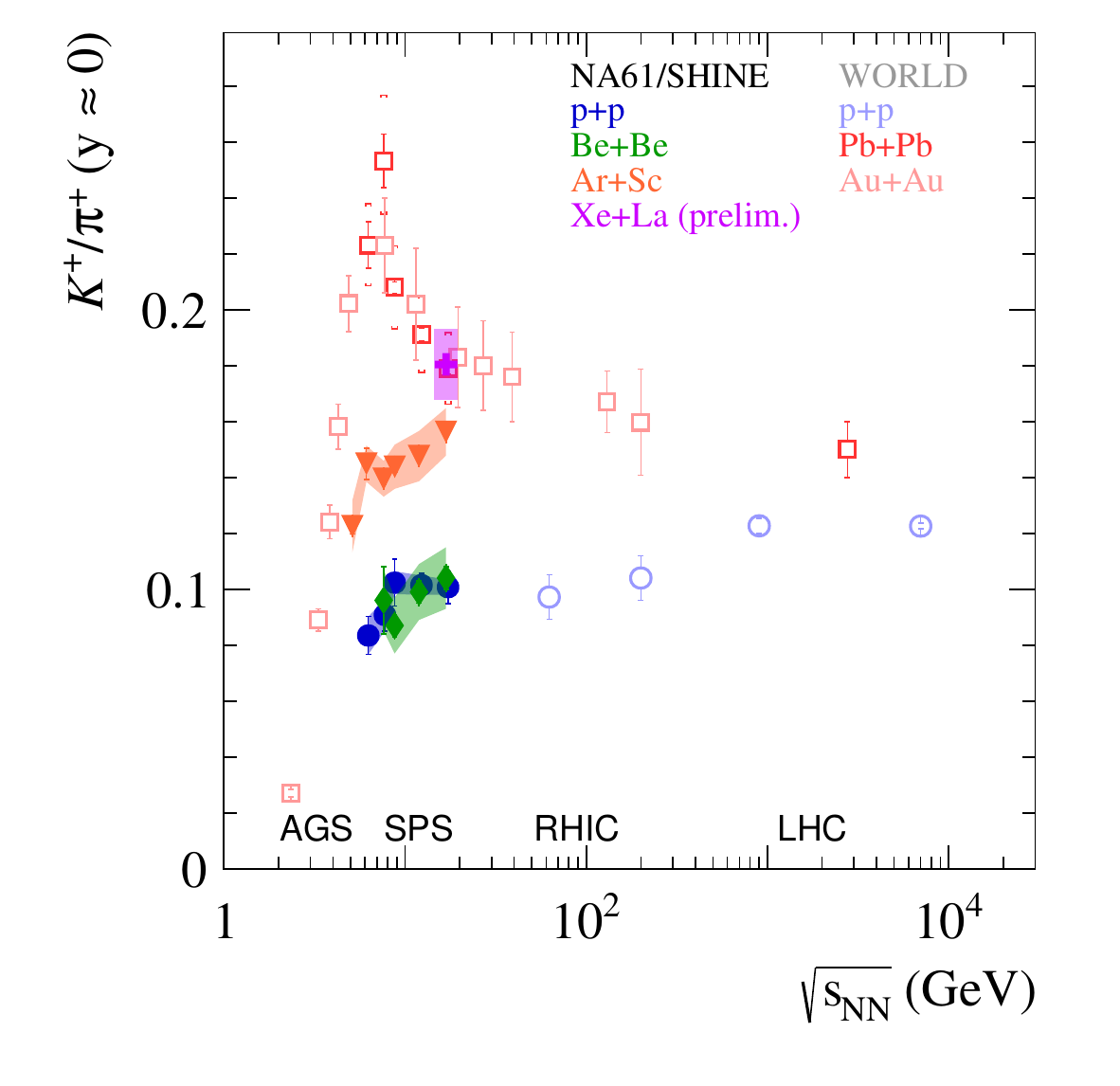}%
  \includegraphics[height=0.198\textheight, page = 1, clip, trim=0 0 0 5mm]{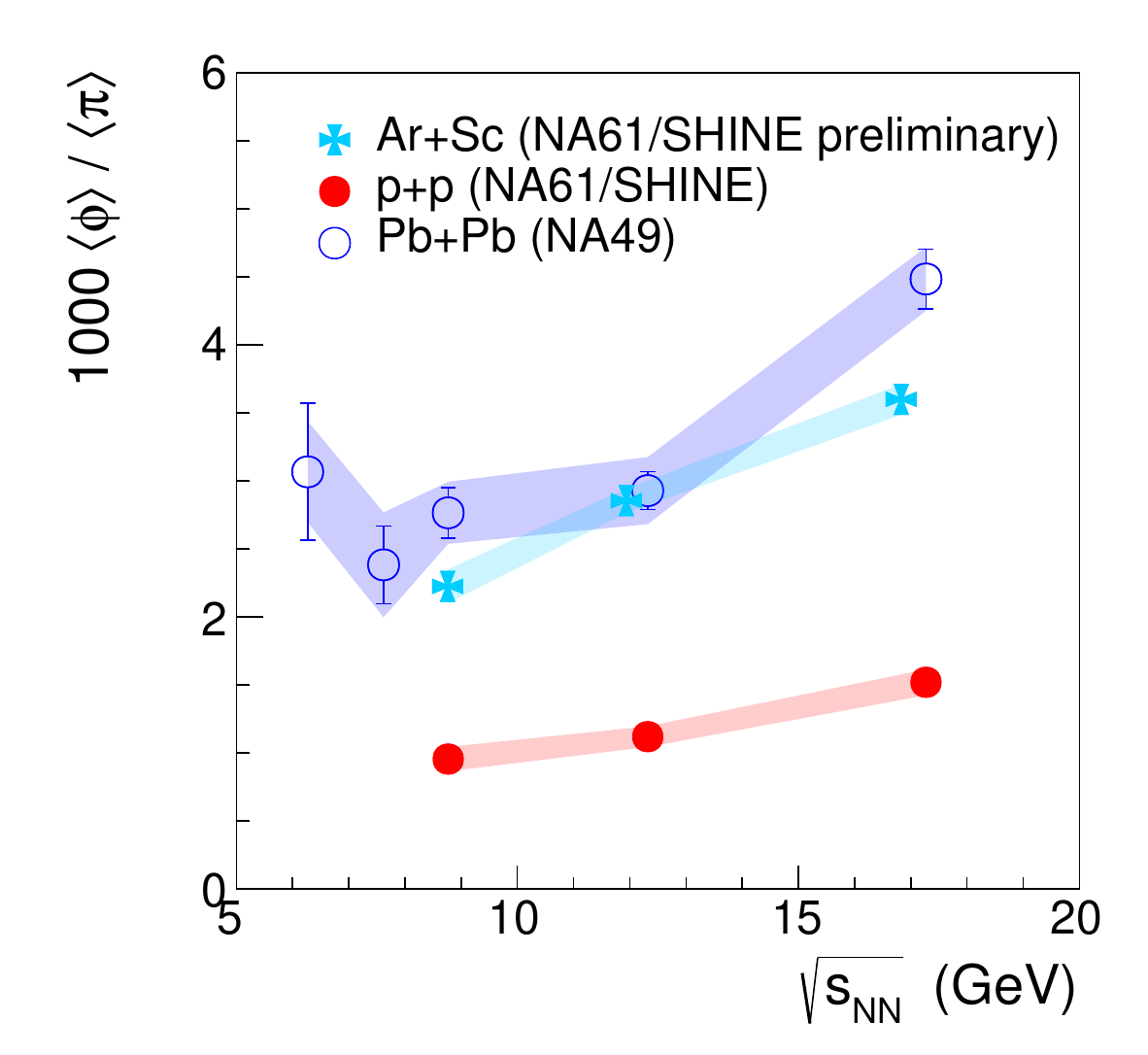}
  \caption{System-size and energy dependence of the $\Kp/\pip$ ratio for 4$\pi$
    yields (\emph{left}) and midrapidity yields (\emph{middle}), and $\phi/\pi$
    ratio for 4$\pi$ yields (\emph{right}) measured by the \NASixtyOne experiment
    and compared to the world data. See
    \protect\recites{NA61SHINE:2023epu,Aduszkiewicz:2019ldi} for
    references to the world data and published \NASixtyOne results.
  }
  \label{fig:onset}
\end{figure}
\begin{figure}[htb]
  \centering
  \hspace*{-1ex}%
  \includegraphics[width=\textwidth,clip, trim=9.6mm 31.5mm 9.6mm 17.3mm]{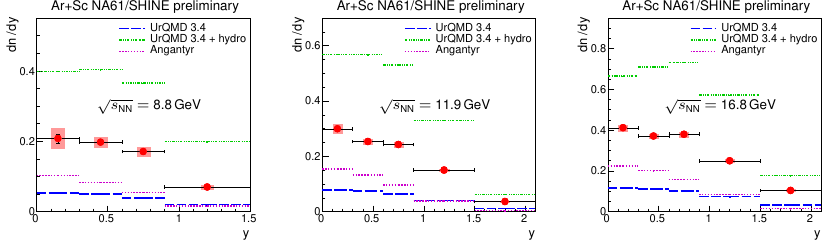}
  \caption{Rapidity distributions of \phiM mesons produced in the 10\% most
    central \ArSc collisions measured by the \NASixtyOne
    experiment (red circles), compared to model predictions. Horizontal solid black
    bars visualize bin sizes, vertical bars show statistical uncertainties, while
    shaded red rectangles systematic uncertainties. Model calculations were
    done by S.~Veli (Technical University of Munich) and T.~Janiec (The
    University of Manchester).}
  \label{fig:phi}
\end{figure}
One of the main observables associated with the study of the transition from
confined to deconfined matter is strangeness production. \Cref{fig:onset},
\emph{left} and \emph{middle} shows
system-size and energy dependence of the $\Kp/\pip$ ratio, connected to
strangeness over entropy ratio. For \PbPb and \AuAu collisions a sharp peak
(\emph{horn}) is visible, explained by the onset of
deconfinement~\cite{Gazdzicki:2010iv}, \ie transition from hadron production
dominated by formation and decays of hadron resonances to that dominated by
creation and hadronization of Quark Gluon Plasma (QGP). A new \NASixtyOne
preliminary \PbPb data point at \sNNn{7.6} confirms NA49 results. System-size
dependence reveals interesting features: the ratio for \BeBe is close to that
for \pp, while \ArSc and new preliminary \XeLa (full energy scan is still
ongoing) data points are higher and approach those for \PbPb at the top SPS
energy. This picture, at the top SPS energy, is consistent with a transition
from hadron production dominated by creation and fragmentation of strings to
that dominated by hadronization of a thermalized fireball of QGP. This
phenomenon, which we call the \emph{onset of fireball}, can be paralleled via
the AdS/CFT duality to creation of black holes~\cite{Kalaydzhyan:2014tfa}.
\par
Complementary to the open strangeness production, \NASixtyOne measures hidden
strangeness production. \Cref{fig:phi} shows rapidity distributions of \phiM
mesons produced in the 10\% most central \ArSc collisions at three energies,
compared to model predictions. These preliminary results are first-ever results
on \phiM production in intermediate-size systems at CERN SPS. It is clear that
none of the considered models matches the experimental data points.
\Cref{fig:onset}, \emph{right} presents enhancement of \phiM production normalized
to pion production. It is evident that the $\phi/\pi$ ratio for \ArSc collisions
is comparable to that in \PbPb reactions, and significantly higher than for \pp
interactions.

\begin{figure}[htb]
  \centering
  \begin{tikzpicture}
    \node (plot) {
      \includegraphics[width=0.4\textwidth]{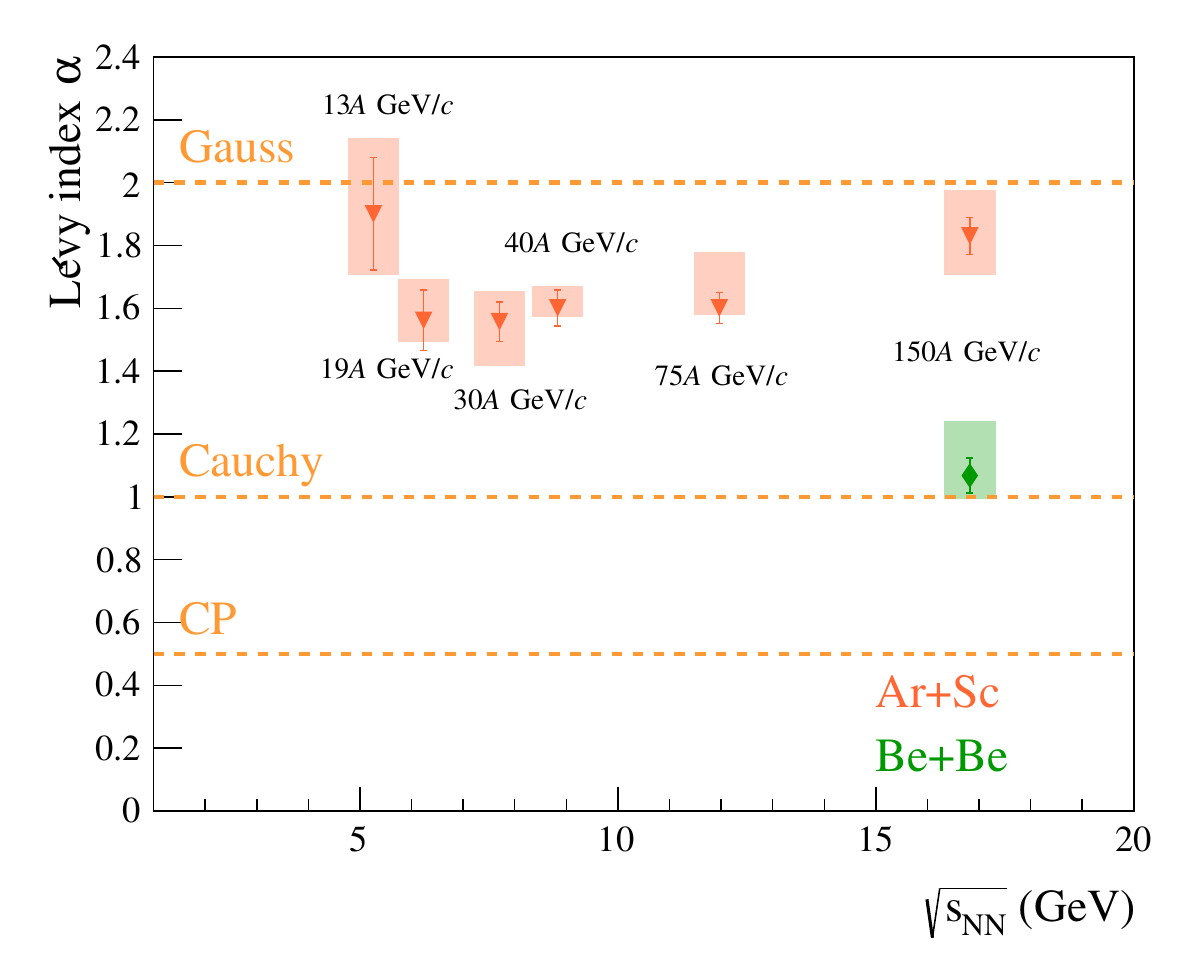}
    };
    \node [anchor=north east, shift=({-1em, -1.5em}), font=\sf\tiny] at (plot.north east) {
      \NASixtyOne preliminary
    };
  \end{tikzpicture}
  \caption{Energy dependence of the preliminary \NASixtyOne femtoscopy results
    for like-sign pions
    in the 10\% most central \ArSc collisions and the published
    result~\protect\cite{NA61SHINE:2023qzr} in the 20\%
    most central \BeBe reactions. Numbers show beam momenta corresponding to the
    points. Vertical bars show statistical uncertainties, while shaded
    rectangles the systematic uncertainties. See the text regarding the
    horizontal dashed lines.}
  \label{fig:femto}
\end{figure}
Among various correlation and fluctuation studies dedicated to the search for
the critical point of strongly interacting matter, \NASixtyOne performed femtoscopy analysis of
like-sign pions in the \BeBe~\cite{NA61SHINE:2023qzr} collisions at the top SPS
energy and recently also in \ArSc reactions at six energies, see
\cref{fig:femto}. L\'evy-type source
assumption was used, yielding Bose-Einstein part of the two-particle correlation function in the form
of $C(q) = 1 + \lambda \cdot e^{-|qR|^\alpha}$, where $q$ is the momentum
difference, $R$ measures the length of homogeneity and the L\'evy stability
parameter $\alpha$ describes the shape of the source. If $\alpha = 2$ the source
is Gaussian, if $\alpha = 1$ it is a Cauchy shape, while 3D Ising model with
random external field predicts $\alpha = 0.5 \pm 0.05$ for the critical
point (CP)~\cite{PhysRevB.52.6659,Csorgo:2005it}. These characteristic values are
indicated in \cref{fig:femto} by horizontal dashed lines. It is clear that
measured $\alpha$ values are far from the critical point prediction, however one
can see potentially interesting non-monotonic structure for \ArSc collisions,
which requires further studies.

\section{Open charm}
\begin{figure}[tb]
  \centering
    \includegraphics[height=0.20\textheight]{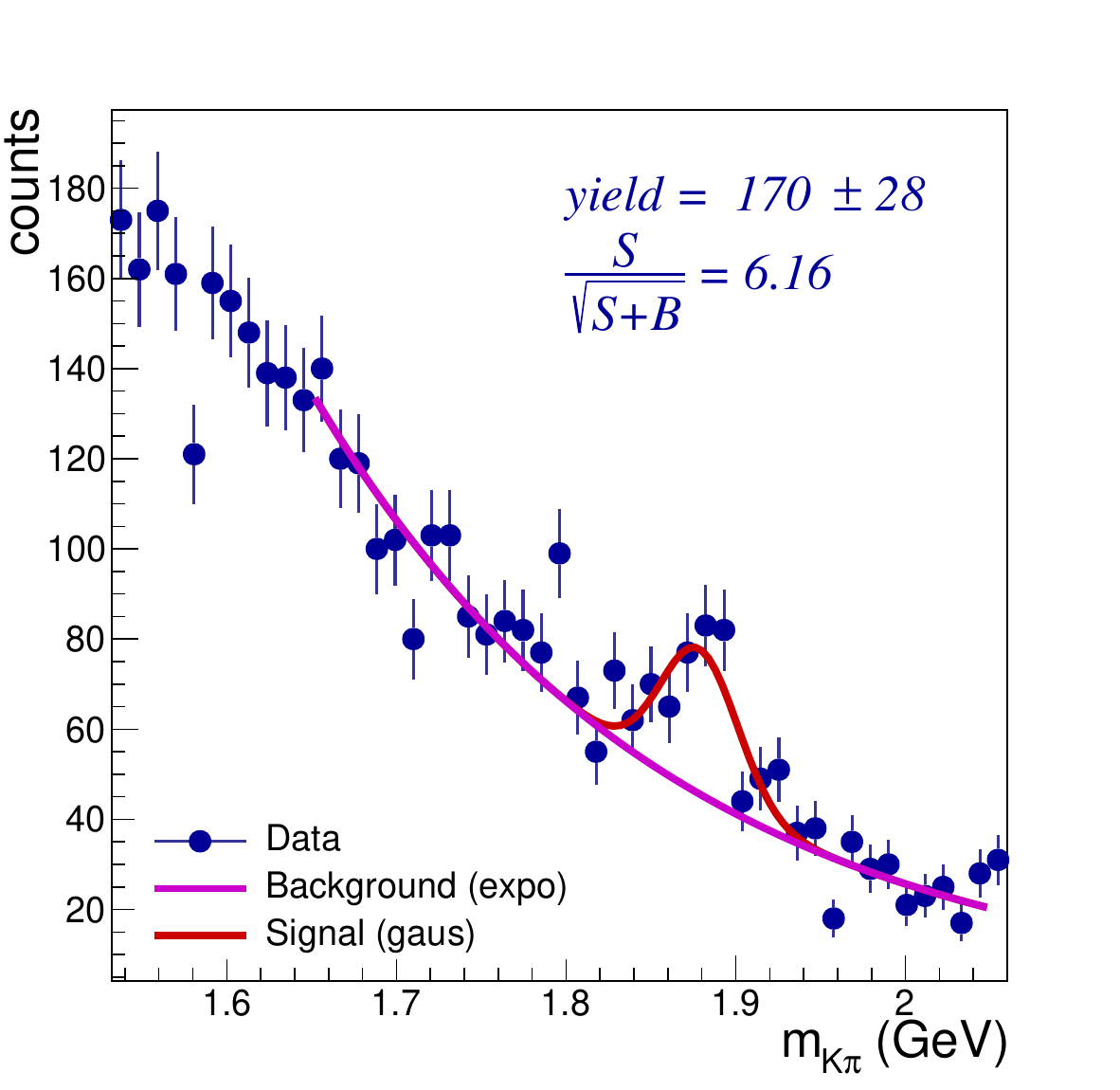}%
    \includegraphics[height=0.20\textheight]{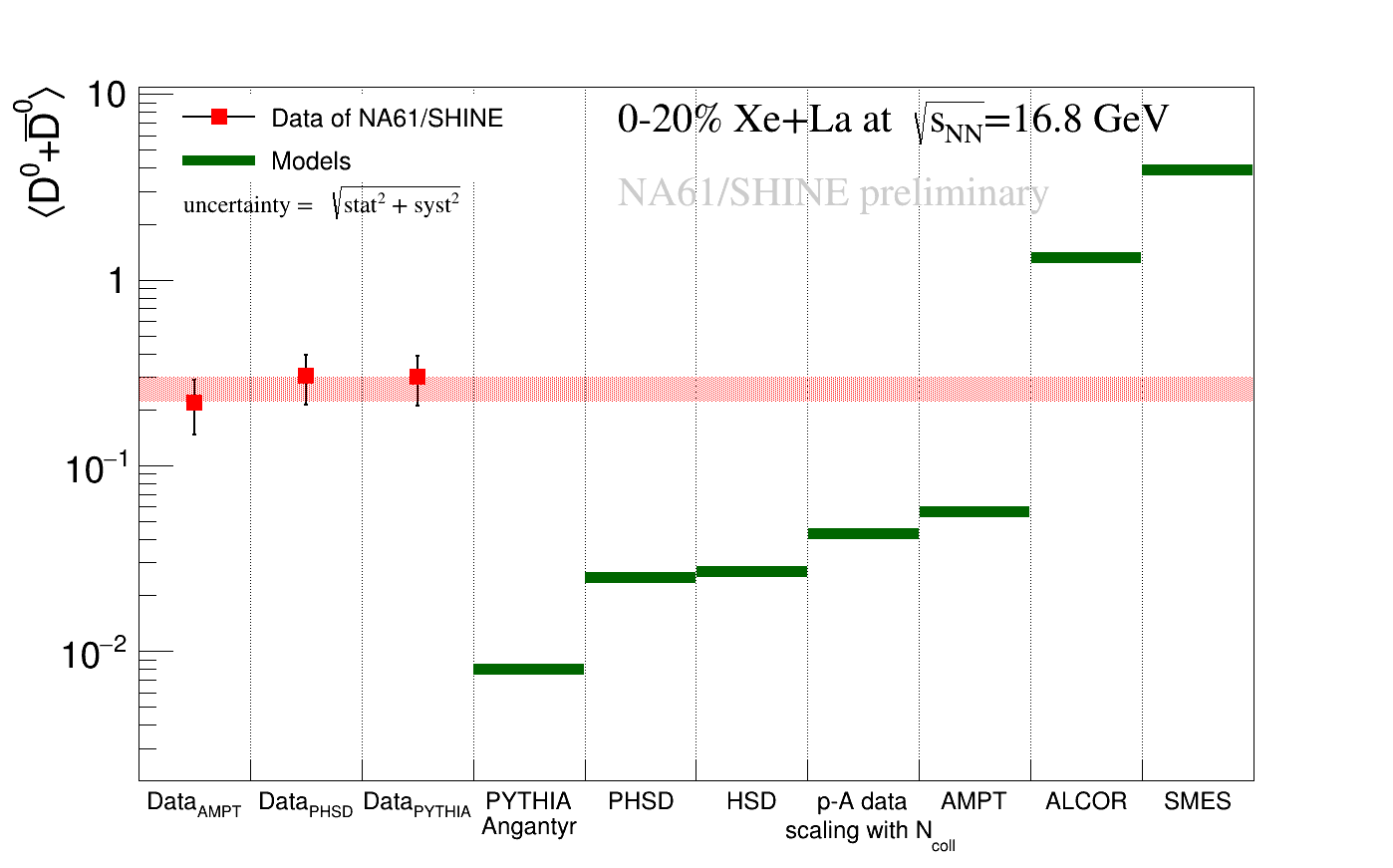}
    \caption{Preliminary \NASixtyOne results on the
      $D^0+\overline{D}^0$ (open charm) 4$\pi$ yield measured in the 20\% most central
      \XeLa collisions at \sNNn{16.8}. \emph{Left:} the invariant mass spectrum
      used to extract the raw value of the yield. \emph{Right:} yield corrected
      for acceptance and efficiency with three different models compared to
      model predictions. The red band indicates uncertainty associated with
      model dependence of the correction.
  }
  \label{fig:charm}
\end{figure}
While the main goal of the post-LS2 physics program of \NASixtyOne is direct
measurement of the open charm production in \PbPb collisions at SPS energies,
the first preliminary results were obtained recently in \XeLa reactions at the
top SPS energy, see \cref{fig:charm}. This is the first-ever direct measurement
of open charm production in nucleus-nucleus collisions at SPS energies.
The data were measured with small acceptance, prototype Vertex
Detector, and had rather small statistics for this analysis.
This allowed only for non-differential measurement in the full phase
space, making the acceptance and efficiency correction model-dependent. Three
different models were used to estimate the correction, giving three different
results shown in \cref{fig:charm}, \emph{right}. As is evident from the plot,
even with relatively large uncertainties, these results put strong
constraints and differentiate between the existing models of hadron production.

\section{Unexpected excess of charged over neutral kaons}
\begin{figure}[tb]
  \centering
  \includegraphics[width=0.44\textwidth]{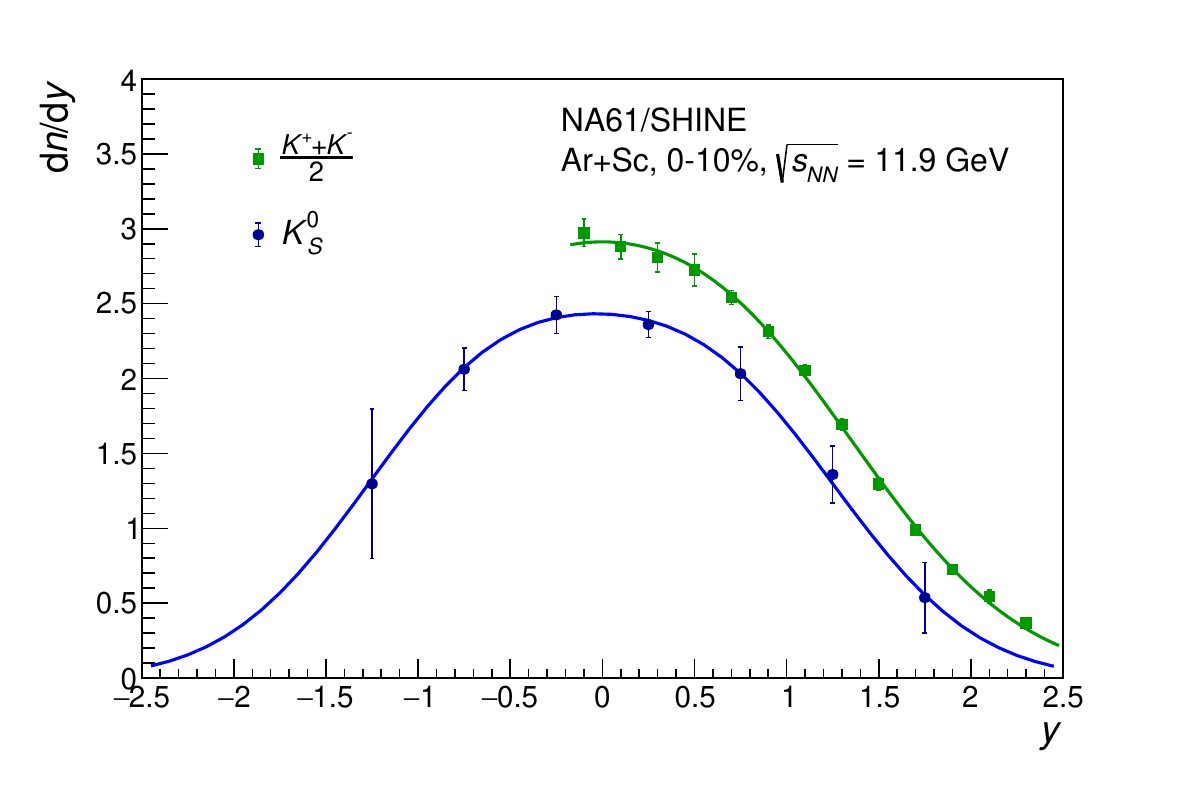}%
  \includegraphics[width=0.56\textwidth]{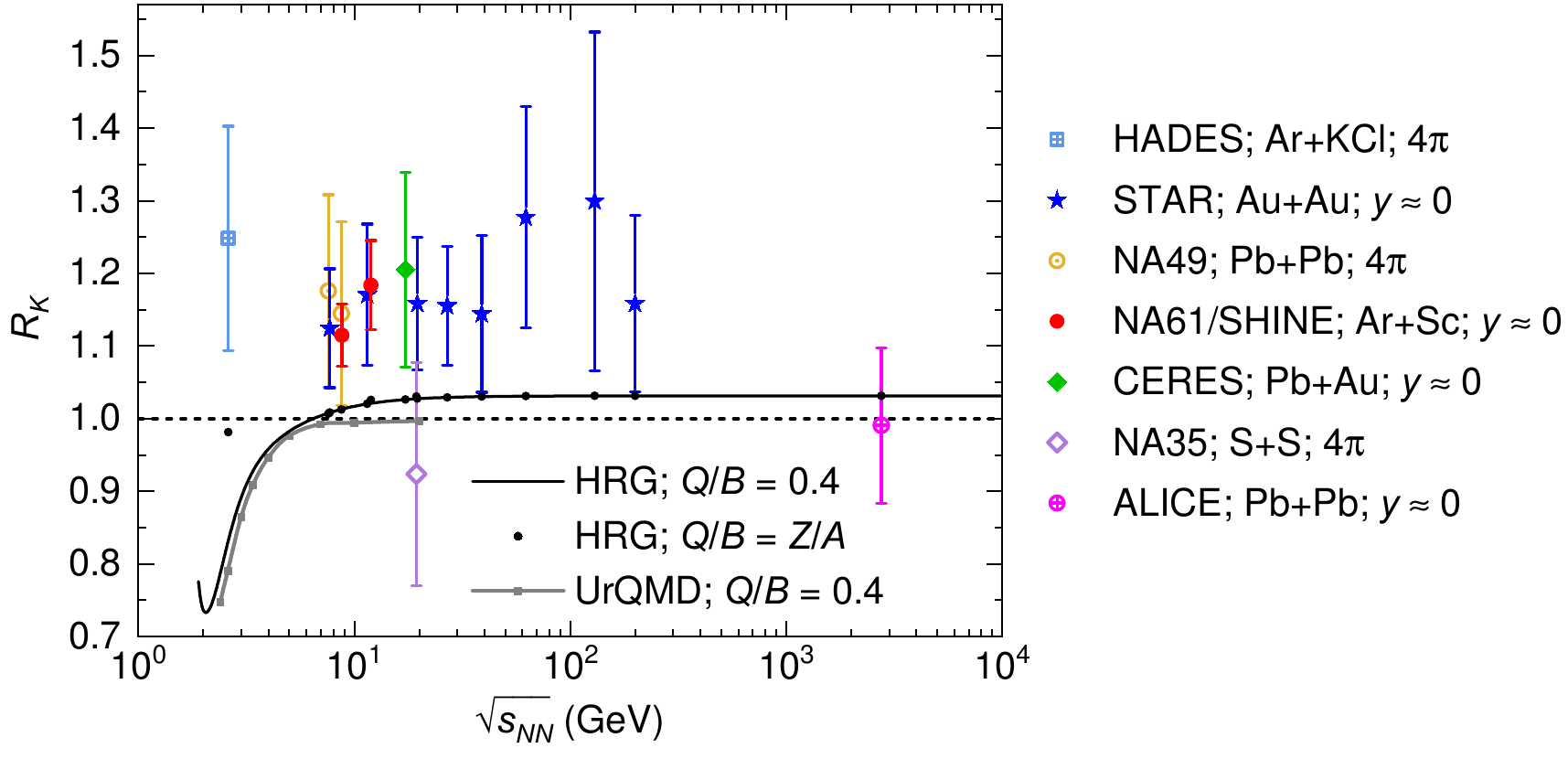}
  \caption{\emph{Left:} comparison~\protect\cite{NA61SHINE:2023azp} of rapidity spectra of neutral to charged
    kaons in the 10\% most central \ArSc collisions at \sNNn{11.9}.
    \emph{Right:} energy dependence of the $R_K$ ratio (see text) measured by
    \NASixtyOne and other experiments~\protect\cite{NA61SHINE:2023azp} and new
    preliminary \NASixtyOne point for \ArSc at \sNNn{8.8}, in 4$\pi$ or midrapidity depending on the
    experiment, compared to model predictions.
  }
  \label{fig:isospin}
\end{figure}
For exact isospin symmetry, \ie equality of $u$ and $d$ quarks and collisions of
charge-symmetric ($Z=N$) nuclei, one expects for the yields of hadrons:
$\expval*{\Kp}(u\bar{s})=\expval*{K^0}(d\bar{s})$
and
$\expval*{\Km}(\bar{u}s)=\expval*{\overline{K}^0}(\bar{d}s)$.
Neglecting small CP violation we also have
$\expval*{K_S^0} = \frac{1}{2}\expval*{K^0} +
    \frac{1}{2}\expval*{\overline{K}^0} = \expval*{K_L^0}$.
Therefore in this case, the expected charged to neutral kaon ratio:
\begin{equation}
  R_K
    = \frac{\expval*{\Kp} + \expval*{\Km}}{\expval*{K^0} + \expval*{\overline{K}^0}}
    = \frac{\expval*{\Kp} + \expval*{\Km}}{2 \expval*{K_S^0}} = 1 \,.
  \label{eq:isospin}
\end{equation}
Surprisingly, \NASixtyOne observed large deviations from unity for this ratio,
first in the \pimC collisions~\cite{NA61SHINE:2022tiz}, then in \ArSc
reactions~\cite{NA61SHINE:2023azp}, see \cref{fig:isospin}, \emph{left}. While
\pimC is not charge-symmetric, so \cref{eq:isospin} is not expected to hold,
the measured value of $R_K$ for \pimC collisions still differs significantly
from that predicted by models as well as quark
counting~\cite{NA61SHINE:2022tiz}. \Cref{fig:isospin}, \emph{right} shows energy
dependence of $R_K$
measured by \NASixtyOne at \sNNn{8.8} and \sNNn{11.9}, together with values
resulting from a compilation of measurements of \Kp, \Km and \KOS from other
experiments performed by the \NASixtyOne Collaboration. The experimental data
points are compared to model predictions.
It is clear that the effect is present in the data from other
experiments, but uncertainties were too large to notice it before the
\NASixtyOne result at \sNNn{11.9} appeared. A new preliminary \ArSc result at
\sNNn{8.8} increases significance of the isospin-symmetry violation beyond the
known effects (based on deviation from Hadron Resonance Gas model) compared to \recite{NA61SHINE:2023azp}
from 4.7$\sigma$ to 5.3$\sigma$. The origin of this effect, in clear
disagreement with theoretical predictions, remains to be elucidated.

\section*{Acknowledgments}
This work was supported by the National Science Centre, Poland (grant number
2023\slash 51\slash D\slash ST2\slash 02950) and the Polish Minister of Education and Science (contract
No. 2021\slash WK\slash 10).

\section*{References}
\bibliography{marcinek}

\begin{thebibliography}{1}

\bibitem{Gazdzicki:2010iv}
M.~Gazdzicki, M.~Gorenstein, and P.~Seyboth.
\newblock {\em Acta Phys. Polon. B}, 42:307--351, 2011.

\bibitem{NA61SHINE:2023epu}
H.~Adhikary et~al.
\newblock {\em Eur. Phys. J. C}, 84(4):416, 2024.

\bibitem{Aduszkiewicz:2019ldi}
A.~Aduszkiewicz et~al.
\newblock {\em Eur.~Phys.~J.~C}, 80(3):199, 2020.

\bibitem{Kalaydzhyan:2014tfa}
T.~Kalaydzhyan and E.~Shuryak.
\newblock {\em Phys. Rev. D}, 90(2):025031, 2014.

\bibitem{NA61SHINE:2023qzr}
H.~Adhikary et~al.
\newblock {\em Eur. Phys. J. C}, 83(10):919, 2023.

\bibitem{PhysRevB.52.6659}
H.~Rieger.
\newblock {\em Phys. Rev. B}, 52:6659--6667, 1995.

\bibitem{Csorgo:2005it}
T.~Csorgo, S.~Hegyi, T.~Novak, and W.~A. Zajc.
\newblock {\em AIP Conf. Proc.}, 828(1):525--532, 2006.

\bibitem{NA61SHINE:2023azp}
H.~Adhikary et~al.
\newblock {\em Nature Commun.}, 16(1):2849, 2025.

\bibitem{NA61SHINE:2022tiz}
H.~Adhikary et~al.
\newblock {\em Phys. Rev. D}, 107(6):062004, 2023.

\end{thebibliography}

\end{document}